\documentclass[a4paper,11pt]{article}
\usepackage{pos}
\usepackage{footmisc}
\usepackage{subcaption}

\title{Progress on the ALETHEIA project and a new approach to mitigate events overlap}

\author[a]{Fabio Acerbi}
\author[b]{Guangpeng An}
\author[c]{Fabrizio Armani}
\author[c]{Giuseppe Di Carlo}
\author[c]{Davide D'Angelo}
\author[a]{Andrea Ficorella}
\author[d]{Yuanning Gao}
\author[a]{Alberto Gola}
\author[b]{Fengbo Gu}
\author[b]{Zihuai Hu}
\author[b]{Zhuo Liang}
\author*[b, e, f]{Junhui Liao}
\author[b]{Edoardo Martinenghi}
\author[b]{Zhaohua Peng}
\author[c]{Valerio Toso}
\author[g]{Lei Zhang}
\author[g]{Lifeng Zhang}
\author[b]{Jian Zheng}

\affiliation[a]{Fondazione Bruno Kessler,
via Sommarive, 18, Trento, 38123, Italy}

\affiliation[b]{Department of nuclear physics, China Institute of Atomic Energy,
No: 1, Sanqiang Rd., Xinzheng, Fangshan district, Beijing, China}

\affiliation[c]{dip. di Fisica Università degli Studi \& I.N.F.N.,
via Celoria, 16 20133 Milano, Italy}

\affiliation[d]{School of Physics, Peking University,
ChengFu Rd. 209, Haidian, 100084, Beijing, China}

\affiliation[e]{Jinping Deep Underground Frontier Science and Dark Matter Key Laboratory of Sichuan Province,
Liangshan, 615000, Sichuan, China}

\affiliation[f]{Yalong River Hydropower Development Company, Ltd,
288 Shuanglin Road, Chengdu, 610051, Sichuan, China}

\affiliation[g]{Department of Nuclear Synthesis Technology,China Institute of Atomic Energy,
No: 1, Sanqiang Rd., Xinzheng, Fangshan district, Beijing, China}

\emailAdd{junhui.private@gmail.com, or junhui\_private@163.com}

\abstract{The ALETHEIA project aims to search for low-mass dark matter using liquid helium (LHe)-filled time projection chambers (TPCs). While liquid argon and liquid xenon TPCs have been extensively employed in the field of direct dark matter detection, successful development of LHe TPCs has not yet been achieved. Launched in 2020, our project has made significant progress since then. These advancements have convinced us that a single-phase LHe TPC is technologically feasible.
Compared to liquid xenon and liquid argon TPCs, one of the unique challenges for LHe TPCs is event overlap caused by the 13-second lifetime scintillation. We will demonstrate that this overlap can be entirely mitigated when the LHe temperature is maintained near 1.0 K. At this temperature, electron mobility is three orders of magnitude higher than at approximately 4.0 K, which is the temperature we initially proposed for the LHe TPC.}

\FullConference{19th International Conference on Topics in Astroparticle and Underground Physics (TAUP2025)\\
24–30 Aug 2025\\
Xichang, China\\}

\begin{document}
\maketitle

\section{ALETHEIA introduction}

In the dark matter (DM) direct detection community, especially in the search for high-mass ($\sim \geq$ 10 GeV/c$^2$) WIMPs, liquid noble gas time projection chambers (TPCs) are arguably the leading technology due to numerous advantages, including but not limited to: (a) extremely low backgrounds within the fiducial volume - primarily caused by the elegant detector design and efficient, continuously recycling purification processes; (b) excellent discrimination between signals (nuclear recoil, NR) and backgrounds (electron recoil, ER), owing to the differing charge densities of NR and ER events; and (c) the ability to understand detector performance in real-time through regular in-situ calibrations. Examples include liquid xenon TPCs like LZ~\cite{LZProject}, PandaX~\cite{PandaXProject}, and XENON~\cite{XenonProject}, as well as liquid argon TPCs such as DarkSide~\cite{DarkSideProject} and DEAP~\cite{DeapProject}. 
To our knowledge, however, no liquid helium (LHe) TPC has yet been constructed. The ALETHEIA project, standing for "A Liquid hElium Time projection cHambEr In dArk matter," aims to carry out the direct detection of low-mass DM and beyond with LHe TPCs~\cite{ALETHEIA2023}.

\section{The progress of ALETHEIA so far}

We launched the ALETHEIA project in 2020 at the China Institute of Atomic Energy (CIAE). Building on the experience accumulated over the past five years, we are confident in our ability to develop a single-phase LHe TPC. (a) We designed and built a custom LHe detector capable of reaching 4 K and maintaining this temperature for several hours~\cite{ALETHEIA2023}. (b) We developed a set of devices to coat tetraphenyl butadiene (TPB) powder onto the inner surfaces of a 10 cm cylindrical chamber\footnote{LHe scintillation peaks at 80 nm (16 eV), which no commercial photosensor can detect directly. A wavelength shifter like TPB is necessary to convert this ultraviolet scintillation light into visible wavelengths suitable for detection.}. Using a scanning electron microscope (SEM), we demonstrated that the TPB coating layer can withstand cooling to 4 K without visible damage~\cite{TBPCoating4K}. (c) Based on measurements at near LHe temperature~\cite{SiPMs4KTests}, we confirmed that the NUV-HD-Cryo type SiPMs from Fondazione Bruno Kessler (FBK) are suitable as photosensors for an LHe TPC. These assessments included analysis of after-pulses, cross-talk, dark count ratio, I-V curves, and photon detection efficiency.

Compared to single-phase TPCs, dual-phase TPCs provide the advantage of three-dimensional event localization - allowing for background rejection through fiducial volume cuts. In an LHe TPC, the event position in X and Y coordinates can be inferred from the pattern of hits on top and bottom photosensors, while the Z position is derived from the drift time of ionized electrons under an external electric field. Important considerations for the drift field include: (a) the fraction of ionized electrons that escape recombination to become free electrons; (b) potential event overlap due to the drift times; and (c) the engineering feasibility of applying and transmitting high voltages into LHe TPCs. We will primarily focus on (a) and (b) here, leaving (c) to be addressed in a dedicated publication.

Measurements with 5.3 MeV alpha particles from a polonium source~\cite{WilliamStacey57} showed that approximately 10\% of ionized electrons escape recombination under a 10 kV/cm drift field, while a 50 kV/cm increases this separation fraction to about 25\%. Our preliminary simulations using COMSOL~\cite{ComsolWebsite} align reasonably well with these experimental results~\cite{ComsolPreliminarySimulations}. After thermalization - occurring within about 10 ps - an electron with initial energy of roughly 7 eV will lose energy down to approximately 1 eV, the threshold at which electrons form "bubbles" in LHe~\cite{bubblesMaris2008}. These electron bubbles have a radius of approximately 2 nm and a mobility as low as $2 \times 10^{-2}$ cm$^2$/(V·s) at 4 K~\cite{LHeProperties}. Under a 10 kV/cm drift field, this corresponds to a drift velocity of about 2 m/s. For a TPC with a height of 1 meter, the drift time is about 0.5 second.

Assuming a background event rate of 1 Hz for an LHe TPC~\footnote{The 1 Hz event rate in an LHe is a reasonable assumption given (a) the LZ background events rate is 40 Hz~\cite{LZTDR} and (b) LHe is much more cleaner than liquid xenon~\cite{ALETHEIA2023}. \label{fn1}}, event overlap would be significant. This is because the average time between consecutive background events is 1 s, which is comparable to the 0.5 s electron drift time required to record a single event. Without a significant increase in electron mobility, an LHe TPC would not achieve desirable performance due to the overlap.
Fortunately, cooling the LHe to 1 K can increase electron mobility by up to three orders of magnitude compared to its value at 4 K ~\cite{LHeProperties}. Consequently, the electron drift time in a 1-meter height LHe TPC at 1 K would be on the millisecond (ms) scale. Since this ~1 ms drift time is three orders shorter than the 1 s interval between background events, event overlap becomes negligible.
While temperatures below 1 K would yield even faster mobility, operating below 0.5 K is not advisable due to the risk of breakdown~\cite{LHeBreakdown}, given that a voltage of several kilovolts will be applied to the helium gas region to generate electroluminescent (S2) signals~\footnote{The mechanism of the breakdown was interpreted in Ref.~\cite{LHeBreakdown}. Given the pressure of a 0.5 K helium gas area is 1 Pa~\footnote{Since the helium gas rides on the LHe, so the helium gas has the same temperature of the LHe, 0.5 K.}, part of the electrons (and ions) generated on the surface of LHe due to the Penning process will enter the 1 Pa vacuum space, then get accelerated under a positive (negative) external field to hit back on the surface of LHe to create more ionizations, some of the newly created electrons (ions) will re-enter the 0.1 Pa space again. The cycles would accumulate more and more electrons (ions) on the liquid surface, and ultimately lead to a breakdown.} . An additional reason to avoid very low temperatures is that the resulting high vacuum (pressure < 0.1 Pa) would significantly reduce the S2 signal yield.
Therefore, the operating temperature of an LHe TPC represents a trade-off between an acceptable event overlap rate and a sufficient S2 signal amplitude, among other factors. Through the R\&D program on the dual-phase detector at CIAE, we aim to determine key parameters for LHe TPCs, including the optimal LHe temperature.

\section{Mitigating the possible events overlap due to the 13 s lifetime scintillation}

In addition to the ALETHEIA project, several other experiments and proposals aim to hunt for dark matter using LHe ($^{3}\mathrm{He}$ or $^{4}\mathrm{He}$) detectors that rely on scintillation light~\cite{HeRALD19, Autti24, TESSERACT25, DeLight25}. Compared to liquid argon and liquid xenon TPCs in the dark matter direct detection community, a unique challenge for these scintillation-based LHe detectors is internal event overlap. This phenomenon refers to the overlap between different signal components within the same event, as opposed to the conventional overlap between two or more distinct events.
Specifically, the 13-second lifetime triplet scintillation (S13) from a single event can overlap with its own other components:
(i) the prompt singlet scintillation (S1) with a 10 ns lifetime,
(ii) the delayed singlet scintillation (S1$'$) with a $\sim$ 1.6 $\mu$s lifetime,
(iii) the $\sim$ 1 ms electroluminescence signals (S2) generated in the helium gas layer by electrons separated under an external field in a TPC, and
(iv) the $\sim$1 ms phonon and roton signals read out by various bolometers.
If an LHe detector could achieve a background rate as low as 1 mHz, the probability of S13 overlapping with other components would be negligible. However, a 1 mHz rate is currently unrealistic, given that the LZ detector has a rate of ~40 Hz~\cite{LZTDR}. As briefly explained in a previous footnote~\ref{fn1}, a background event rate of 1 Hz is a more reasonable assumption for LHe. Since our 1 K LHe TPC will not use bolometers to read out phonons and rotons, we will exclude these signals from the following analysis.

In addition to our ALETHEIA project, there are a number of experiments/proposals aiming to hunt for dark matter with LHe ($^{3}\mathrm{He}$ or $^{4}\mathrm{He}$) detectors relying on the scintillations emitted from LHe~\cite{HeRALD19, Autti24, TESSERACT25, DeLight25}. Compared to the liquid argon and the liquid xenon TPCs in the community of dark matter direct detection, one of the unique challenges for an LHe detector (employed the scintillations) is the event's internal overlap, which refers to the overlap happens between the different components of the same event (not the conventional one corresponding to the overlap between two or more different events). Specifically, the 13 second lifetime triplet scintillation (S13) of an event can overlap with other components of the same event: (i) the 10 ns lifetime prompt singlet scintillation, S1, (ii) the $\sim$ 1.6 $\mu$s lifetime delayed singlet scintillation, S1$'$, (iii) the $\sim$ 1 ms electroluminescence signals generated in the helium gas layer by the separated electrons under an external field in a TPC, S2, and (iv) the $\sim$ 1 ms phonons and rotons signals read with variant bolometers. If an LHe detector can reach a background rate as low as 1 mHz, the chance of S13 overlap with the other components is ignorable, however, a 1 mHz rate sounds unrealistic for the moment given the LZ detector has the rate of $\sim$40 Hz~\cite{LZTDR}. As briefly explained in footnote~\ref{fn1}, the background events rate in an LHe can be reasonably assumed to be 1 Hz. Since there will be no bolometers to read out phonons/rotons in our 1 K LHe TPC, we will skip the phonons/rotons signals in the following analysis.

\begin{figure}	
	\centering
	\begin{subfigure}[t]{2.5in}
		\centering
		\includegraphics[width=1.1\textwidth, height=0.7\textwidth]{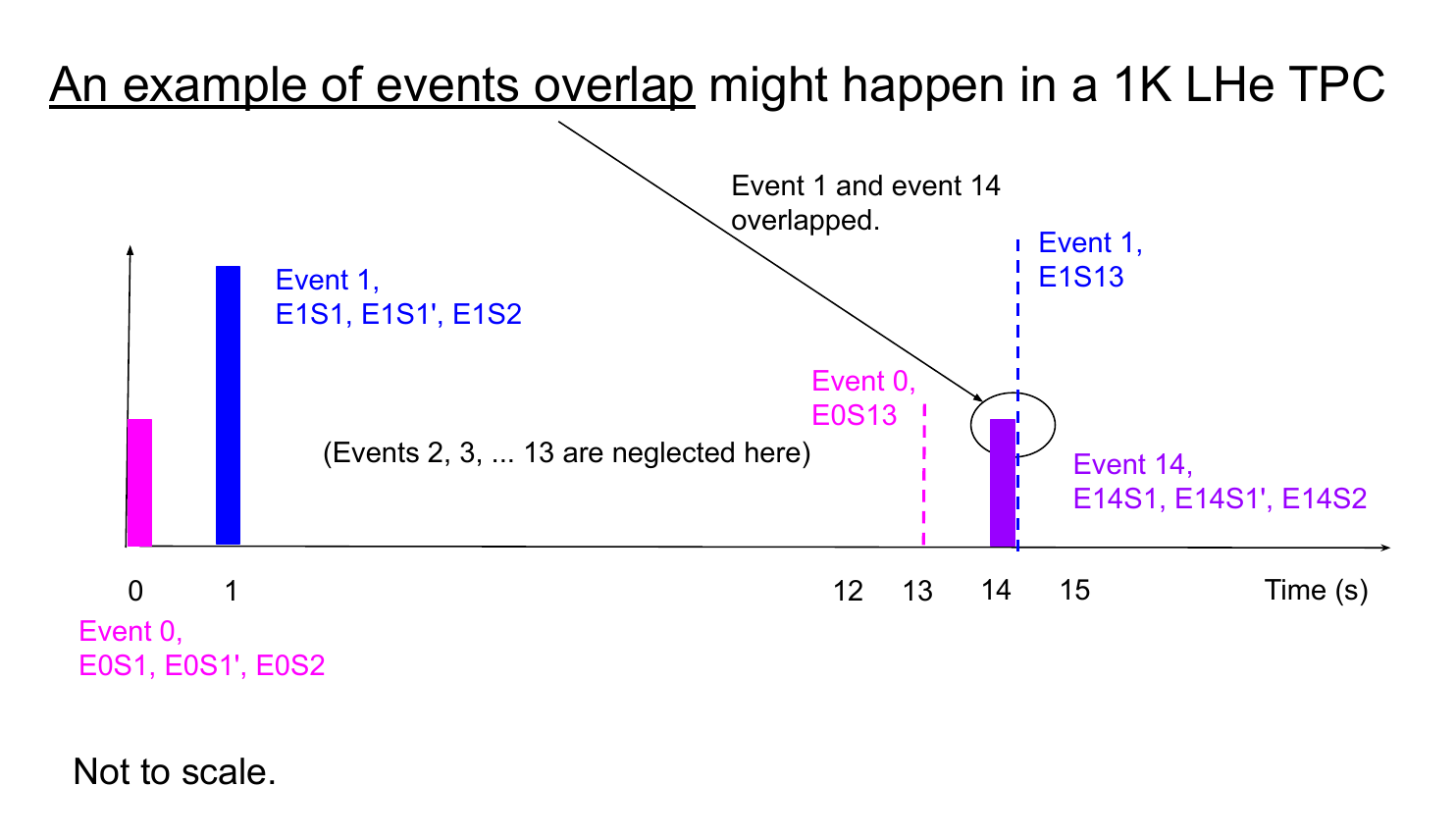}
		\caption{Fig.~\ref{LHeDetectorsEventsOverlap.a}. An example of event overlap in an LHe TPC is shown, where the 13-second lifetime scintillation from Event 1 (E1S13, represented by the long, broken blue vertical line) overlaps with the components of Event 14 (E14S1, E14S1$'$, and E14S2, shown as the short, solid purple vertical bar.} \label{LHeDetectorsEventsOverlap.a}	
	\end{subfigure}
	\quad
	\begin{subfigure}[t]{2.5in}
		\centering
		\includegraphics[width=1.1\textwidth, height=0.7\textwidth]{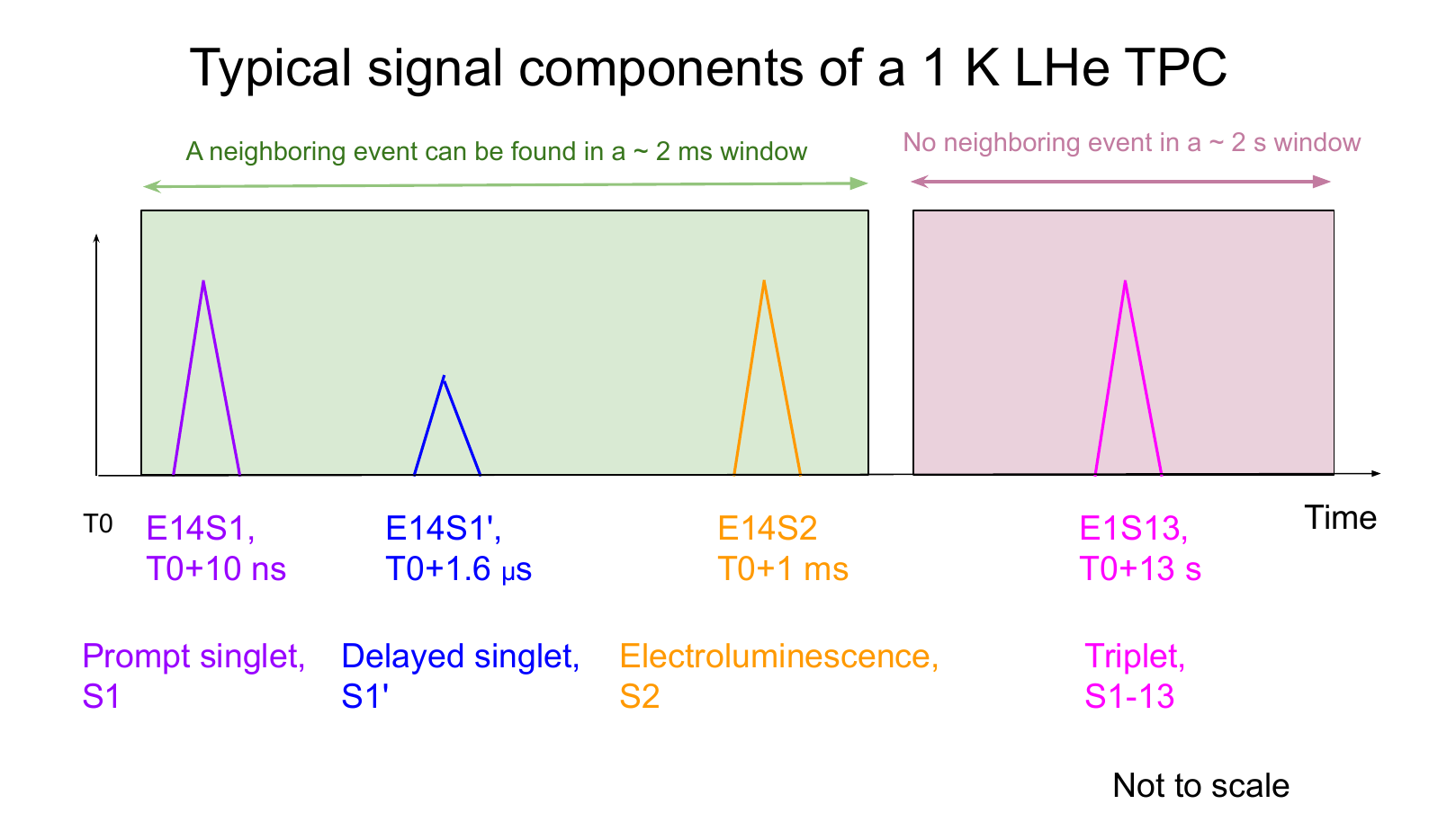}
		\caption{Fig.~\ref{LHeDetectorsEventsOverlap.b}. The overlap can be effectively mitigated in a 1 K LHe TPC. A signal from S1, S1$'$, or S2 will have a neighboring signal within a $\pm$ 1 ms window, whereas a S13 signal will have no neighboring signal within a 2 ($\pm$ 1) s window, assuming a 1 Hz background rate.} \label{LHeDetectorsEventsOverlap.b}
	\end{subfigure}
	\caption{An example of event overlap arising from the 13 s scintillation lifetime is shown, along with its significant mitigation in a 1 K LHe TPC under the assumption of a 1 Hz background event rate.}\label{LHeDetectorsEventsOverlap}
\end{figure}

Figure~\ref{LHeDetectorsEventsOverlap.a} provides a schematic of event overlap for two events separated by 13 seconds, assuming the 1 Hz background rate discussed previously. The diagram lists 14 events in total. Each event consists of two parts: Part 1 includes the relatively fast components (S1, S1$'$, and S2), denoted for Event 0 as E0S1, E0S1$'$, and E0S2, respectively. Part 2 is the S13 signal (e.g., E0S13 for Event 0). In this scenario, overlap occurs between Event 14's Part 1 (E14S1, E14S1$'$, and E14S2) and Event 1's Part 2 signal (E1S13).

Figure~\ref{LHeDetectorsEventsOverlap.b} demonstrates how this overlap can be entirely mitigated by the following algorithm. For any registered signal, the system searches for neighboring signals within a $\pm$ 1 ms time window (a 2 ms total window). If neighboring signals are present, they are all classified as belonging to Part 1 of the same event and can be further identified as S1, S1$'$, or S2 based on their timing sequence. If no signals are found in the window, the registered signal must be an S13 scintillation. This is because a triplet scintillation is unlikely to have any neighboring signals within a much wider window of $\sim$ 2 seconds, given the hypothesized 1 Hz background rate.
The analysis above assumed a 1 Hz background rate. However, the discriminating capability of a 1 K LHe TPC remains effective even at a higher rate of 40 Hz, comparable to the LZ experiment. The reason is that for an S13 signal, the time window for seeing a neighbor would be 50 ms, which is 25 times wider than the 2 ms window affiliated to Part 1 signals. This large difference in timescales preserves the algorithm's ability to distinguish between signal components.

\captionsetup[subfigure]{labelformat=empty}

\section{Summary}
Based on the experience accumulated over the past five years, we are more confident than ever in our ability to successfully build a single-phase LHe TPC. To fully exploit the physics potential of LHe, we are now developing a dual-phase TPC. The internal event overlap caused by the 13-second triplet scintillation is a challenge unique to LHe detectors. By cooling the LHe to 1 K, this overlap can be entirely mitigated.

\section{Acknowledgement}
Junhui Liao would like to thank Prof. Yong-Hamb Kim at the Institute for Basic Science in Daejeon, South Korea for useful discussions on liquid helium physics. This work has been supported by NSFC (National Natural Science Foundation of China) under the contract of 12ED232612001001 and the ``Continuous-Support Basic Scientific Research Project'' in China.

\end{document}